\documentclass[prl,twocolumn,superscriptaddress]{revtex4-1}
\usepackage[utf8]{inputenc}
\usepackage[american,british]{babel}
\usepackage[T1]{fontenc}
\usepackage[pdftex]{graphicx}  
\usepackage{graphicx, xcolor}
\usepackage{dcolumn}
\usepackage{bm}
\usepackage{amsmath,amsthm,amssymb}
\usepackage{color}
\usepackage{verbatim}
\usepackage{ulem}

\definecolor{darkGreen}{RGB}{0,110,0}
\definecolor{darkBlue}{RGB}{0,0,130}
\usepackage[colorlinks,citecolor=darkGreen,linkcolor=darkBlue,urlcolor=blue,hyperindex]{hyperref}

\newcommand{\lsim}{\raisebox{-0.13cm}{~\shortstack{$<$ \\[-0.07cm]
      $\sim$}}~}
\newcommand{\bra}[1]{\left\langle #1 \right|}
\newcommand{\ket}[1]{\left| #1 \right\rangle}
\newcommand{\braket}[2]{\left\langle #1 \middle| #2 \right\rangle}

\begin{document}
\title{Dynamical preparation of quantum spin liquids in Rydberg atom arrays}
\author{G. Giudici}
\affiliation{Institute for Theoretical Physics, University of Innsbruck, Innsbruck A-6020, Austria}
\affiliation{Institute for Quantum Optics and Quantum Information,Austrian Academy of Sciences, Innsbruck A-6020, Austria}
\affiliation{Munich Center for Quantum Science and Technology (MCQST), Schellingstra\ss{}e~4, D-80799 M\"{u}nchen, Germany}
\affiliation{Arnold Sommerfeld Center for Theoretical Physics, University of Munich, Theresienstr. 37, 80333 M\"{u}nchen, Germany}
\author{M. D. Lukin}
\affiliation{Department of Physics, Harvard University, Cambridge, Massachusetts 02138, USA}
\author{H. Pichler}
\affiliation{Institute for Theoretical Physics, University of Innsbruck, Innsbruck A-6020, Austria}
\affiliation{Institute for Quantum Optics and Quantum Information,Austrian Academy of Sciences, Innsbruck A-6020, Austria}

\date{\today}

\begin{abstract}
We theoretically analyze recent experiments [G.~Semeghini et al., Science 374, 1242 (2021)] demonstrating the onset of a topological spin liquid using a programmable quantum simulator based on Rydberg atom arrays. 
In the experiment, robust signatures of topological order emerge in out-of-equilibrium states that are prepared using a quasi-adiabatic state preparation protocol. 
We show theoretically that the state preparation protocol can be optimized to target the fixed point of the topological phase -- the resonating valence bond (RVB) state of hard dimers -- in a time that scales linearly with the number of atoms.
Moreover, we provide a two-parameter variational manifold of tensor network (TN) states that accurately describe the many-body dynamics of the preparation process. Using this approach we analyze the nature of the non-equilibrium state, establishing the emergence of topological order. 
\end{abstract} 

\maketitle

\begin{figure}
\hspace*{-0.3cm}
 \includegraphics[scale=0.5]{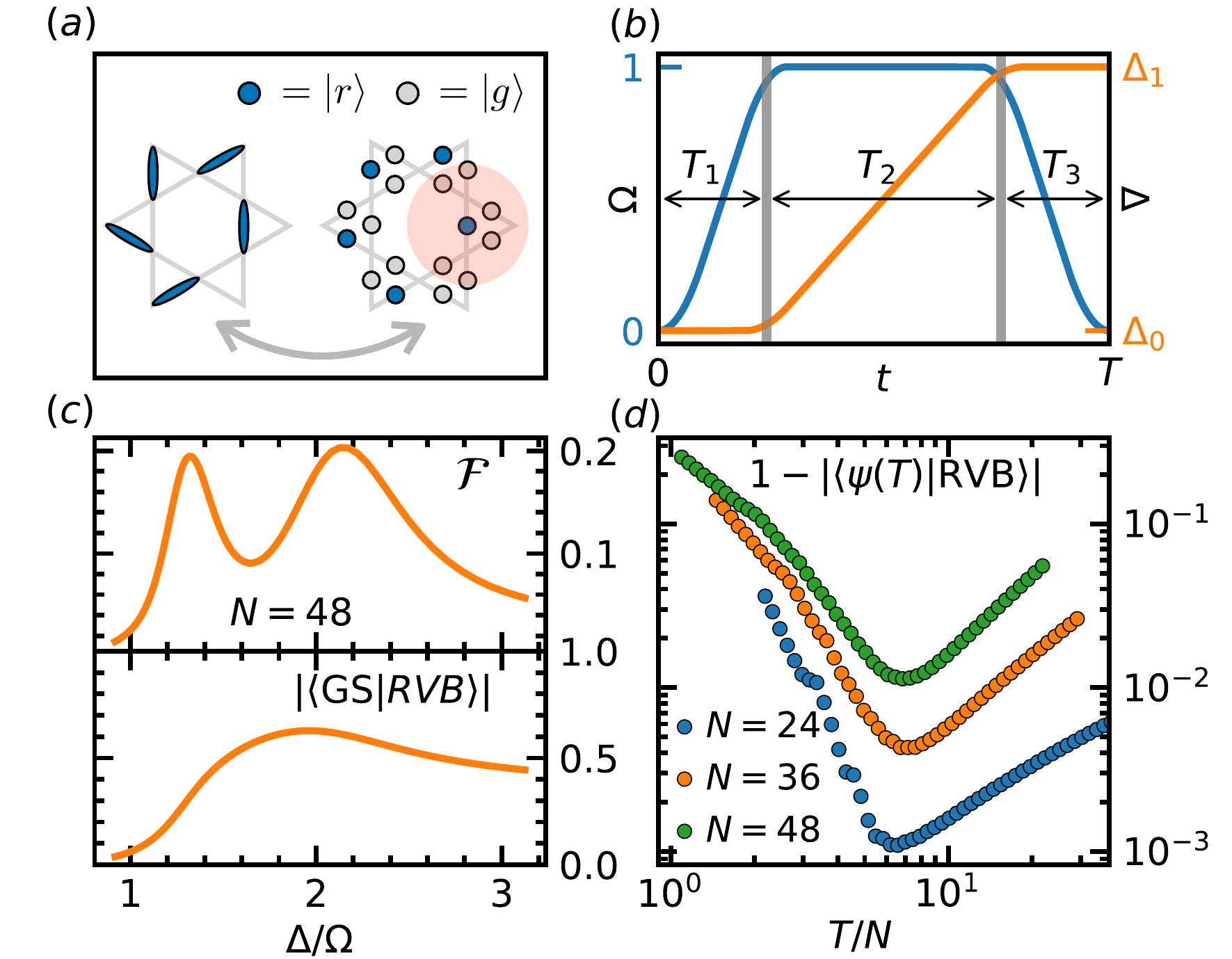}
\caption{(a) Mapping between a dimer model on the kagome lattice and the Rydberg atoms system on the ruby lattice (medial lattice of the kagome). When $R_b \gtrsim 2$ (shaded red circle) the Rydberg constraint is equivalent to the dimer constraint. (b) Schematic representation of the adiabatic state preparation protocol.
(c) Top: Groundstate fidelity susceptibility $\mathcal{F} = (1 - |\braket{\mathrm{GS}(\lambda)}{\mathrm{GS}(\lambda+d \lambda)}|)/d \lambda$ with $\lambda = \Omega/\Delta$ and $d \lambda = 0.0025$. Bottom: RVB overlap $|\braket{\mathrm{RVB}}{\mathrm{GS}}|$ for the groundstate of the Hamiltonian Eq.~\eqref{Ham_constr}. (d) Overlap between the dynamically prepared state and the RVB state as a function of the total sweep time $T$ rescaled with the number of atoms $N$, for $\Delta_0 = -5, \Delta_1 = 1.5$, $T_1 = T_3 = 0.1 T$ and $T_2 = 0.8 T$. }
 \label{fig1}
\end{figure}

\paragraph{Introduction. --} 
Quantum spin liquids (QSLs) arise from the competition between classical frustration and quantum fluctuations \cite{Read1991,Wen1991,Sachdev1992,Savary2016,Wen2017,Sachdev2018}. They are paradigmatic examples of topological quantum matter \cite{Wen2017,Sachdev2018}, characterized by long-range entanglement \cite{Kitaev2003}, hidden non-local order \cite{Wen2017}, and exotic excitations~\cite{Wilczek1982}. Experimental realization and control of topological matter is of central importance not only for understanding 
these many-body quantum phenomena but also for the realization of novel approaches to fault-tolerant topological quantum computation~\cite{Kitaev2003,Nayak2008}. Recently, the onset of a topological spin liquid has been observed in a quantum simulator based on Rydberg atom arrays~\cite{Semeghini2021}.  Following a theoretical prediction \cite{Verresen2021},  the key idea is to exploit the Rydberg blockade mechanism~\cite{jakschFastQuantumGates2000b,lukinDipoleBlockadeQuantum2001,urbanObservationRydbergBlockade2009,gaetanObservationCollectiveExcitation2009} to realize a dimer model, where spin liquid states are known to emerge as equilibrium states at zero temperature~\cite{Rokshar1988,Moessner2001,Moessner2011}. 
These states share many similarities with a resonating valence bond (RVB) state~\cite{Anderson1973} of hard dimers, where the role of a dimer is played by an excited Rydberg state on the medial lattice of a kagome lattice [Fig.~\ref{fig1}a]. While the RVB state is an equal weight superposition of all the (exponentially many) maximal dimer coverings of the kagome lattice, the  Rydberg array can accommodate defects, e.g., in the form of uncovered kagome vertices. Notably, theoretical analysis showed that the presence of a topological phase depends delicately on the precise details of  the Rydberg interactions and atomic positions~\cite{Verresen2021}. Remarkably, experiments showed that robust signatures of quantum spin liquids appear using quasi-adiabatic detuning sweep employed in Ref.~\cite{Semeghini2021}, even in regimes where QSLs are not expected to be stable as the ground state. 
Understanding the dynamical preparation process, robustness of the emerging state, the role of the defects, and the extent to which they can be reduced is crucial for determining the physical properties of the non-equilibrium state as well as its potential utility for topological quantum information processing.\\
\indent
In this Letter we investigate the state produced through the quasi-adiabatic sweep by simulating the quantum dynamics via exact and variational methods. We show that the defect-free RVB state can be prepared with high fidelity in a time that scales linearly with the number of atoms. To understand the nature of the defects generated during the state preparation protocol utilized in the experiments ~\cite{Semeghini2021}, we introduce a novel tensor network  (TN) ansatz for the many-body states.
We demonstrate that this simple ansatz accurately describes the entire many-body dynamics of the preparation process, and we analyze the resulting phase diagram via TN techniques. The latter allows us to study the properties of the non-equilibrium state on system sizes comparable to what can be realized in experiments~\cite{Semeghini2021}. 
By computing several witnesses including non-local order parameters~\cite{bricmont1983,fredenhagen1983} and topological entanglement entropy~\cite{Kitaev2006,levin2006}, we establish the presence of an extended region in parameter space that is adiabatically connected to the RVB state and hosts topological order.

\paragraph{Model Hamiltonian and RVB state preparation. --} 
The Rydberg atom quantum simulator of Ref.~\cite{Semeghini2021} consists of neutral atoms optically trapped in fixed positions on the links of a kagome lattice. Optical transitions between the groundstate $\ket{g}$ and the excited Rydberg state $\ket{r}$ of each atom are controlled via a two-photon process with Rabi frequency $\Omega$ and detuning $\Delta$. Excited Rydberg states interact through Van der Waals potential. The effective Hamiltonian is~\cite{RevModPhys.82.2313,browaeysManybodyPhysicsIndividually2020}
\begin{equation}
\label{Ham_constr}
H = \frac{\Omega}{2} \sum_i \sigma_i^x - \Delta \sum_i n_i +  V \sum_{i>j} \frac{n_i n_j}{|i-j|^6} \quad ,
\end{equation}
where $\sigma_i^x = \ket{g}_i\!\bra{r} + \ket{r}_i\!\bra{g} $ and $n_i = \ket{r}_i\!\bra{r}$.
The parameter $V$ is tuned by varying the lattice spacing, and its magnitude determines the blockade radius from $R_b = (V/\Omega)^{1/6}$. The interactions effectively suppresses simultaneous occupancy of excited Rydberg states for atoms at distance $r\leq R_b$. Numerical calculations are performed enforcing this constraint exactly on periodic clusters at $R_b = 2 a$, where $a$, consistently with Ref.~\cite{Semeghini2021}, is the minimum distance between the atoms. Moreover, we neglect the longer-range tails of the Van der Waals interactons at $r>R_b$ [Fig.~\ref{fig1}a]. The effect of the tails and of a relaxed blockade constraint on the many-body dynamics is described in the final part of the Letter.
The phase diagram of the simplified model hosts three phases: trivially disordered, topologically ordered, and trivially ordered as $\Delta$ increases~\cite{Verresen2021}. These three phases can be identified from the exact diagonalization calculations plotted in Fig.~\ref{fig1}c. The upper panel shows two clear peaks in the groundstate fidelity susceptibility $\mathcal{F} = (1 - |\braket{\mathrm{GS}(\lambda)}{\mathrm{GS}(\lambda+d \lambda)}|)/d \lambda$ that signal an intermediate phase, characterized by high overlap ($\simeq 0.7$ for $N=48$ atoms) with the RVB state [Fig.~\ref{fig1}c, bottom].\\
\indent We first focus on state preparation protocols of the type depicted in Fig.~\ref{fig1}b. The initial state is the vacuum state, where all the atoms are in their ground states. The driving field is turned on at fixed detuning $\Delta_0$ and $\Omega$ increases until it reaches its maximum value (which sets our unit of energy and time). The detuning is then increased from $\Delta_0$ to $\Delta_1$. Finally, $\Omega$ is switched off at fixed detuning $\Delta_1$. The durations of the three stages of the sweep are $T_1,T_2,T_3$ respectively, and the total time is $T$. These parameters can be tuned at will, and we choose $T_1 = T_3 = 0.1 T$, $T_2 = 0.8 T$, $\Delta_0 = -5$ and $\Delta_1 = 1.5$ in units of the maximum Rabi frequency. The final result moderately depends on the parameter $\Delta_1$, while it is mildly affected by the others~\footnote{see the Supplementary Material}.
The parameter sweep is reminiscent of an adiabatic state preparation protocol, in which the instantaneous groundstate runs through all three distinct phases and terminates in the VBS phase~\footnote{Related, but reverse sweeps have been considered in \cite{chandranKibbleZurekScaling2013}.}. However, we are mostly interested in evolution times $T$ that are shorter than those required for complete adiabaticity.  Fig.~\ref{fig1}d shows the overlap of the final state with the defect-free RVB state as a function of the total time $T$. The large and small $T$ regimes are characterized by small overlap with the RVB state. In the former, one recovers the adiabatic limit, where the final state is a valence bond crystal, i.e. the ground state at large detuning~\cite{Verresen2021}. In the latter, the high sweep rate creates a high density of defects on top of the maximal density subspace. Remarkably, at intermediate $T$ the prepared state reaches $0.99$ fidelity with the RVB state for $N=48$ atoms. The maximal overlap with the RVB state is obtained at a time $T^\ast$ that scales linearly with system size $N$, as it is evident from Fig.~\ref{fig1}d, where the overlap is plotted as a function $T/N$. We note that the overlap of this dynamically prepared state and the RVB state exceeds the one between the RVB state and any groundstate of the Hamiltonian by almost two orders of magnitude [Fig.~\ref{fig1}(c,d)].
\begin{figure}
 \hspace*{-0.2cm}
 \includegraphics[scale=0.5]{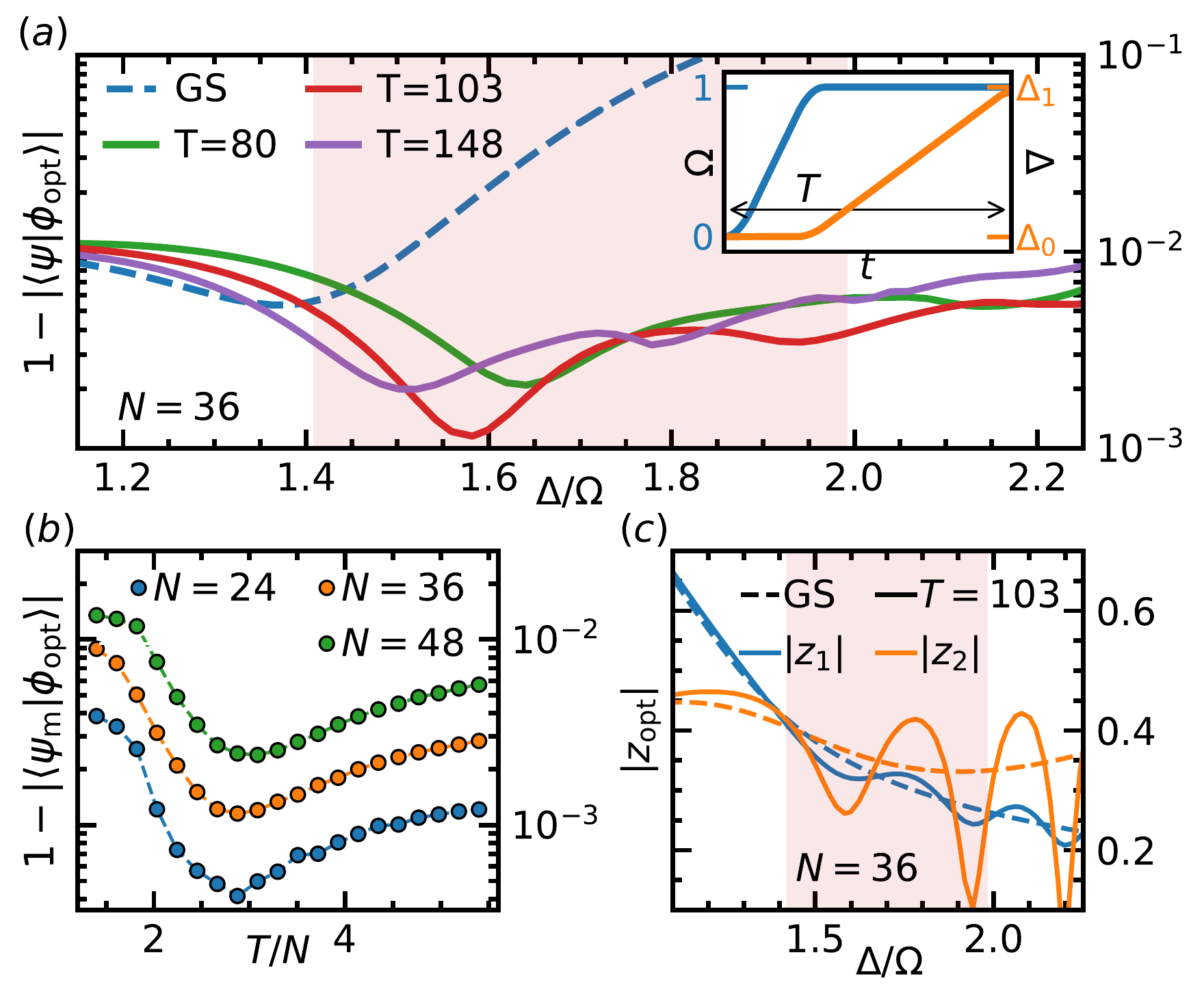}
\caption{(a) Overlap between the dynamically prepared state in the middle of the sweep [depicted in the inset] and the ansatz Eq.~\eqref{ansatz}, optimized over the variational parameters $z_1$ and $z_2$ for various total sweep times $T$, for a periodic cluster of $N=36$ atoms. The dashed line is the optimized overlap with the instantaneous groundstate ($T=\infty$). The shaded red region delimits the topological phase in the groundstate. (b) Maximal overlap obtained during the sweep when $1 \lesssim \Delta/\Omega \lesssim 2 $, for different system sizes $N$, as a function of the total sweep time rescaled by the number of atoms. (c) Optimal absolute values for the variational parameters $z_1$ and $z_2$ in the ansatz state Eq.~\eqref{ansatz} during the sweep. The solid line is the semi-adiabatic sweep that yields the largest overlap ($T=103$ for $N=36$), dashed line is the fully adiabatic sweep ($T=\infty$).  }
\label{fig2}
\end{figure}
\paragraph{Ansatz for the preparation dynamics. --} 
We now expand our focus beyond the analysis of the final state at the end of the sweep, and aim at developing an understanding of the dynamics of the system during the entire state preparation protocol. We are particularly interested in regimes where the dynamics is not adiabatic and the resulting density of monomers is not vanishing small. 
For this we find it convenient to slightly modify the state preparation protocol, to the one depicted in the inset of Fig.~\ref{fig2}a: after the initial switching on of the Rabi frequency, the detuning is linearly increased until the end of the process, i.e., we set $T_3$ to zero [cf.~Fig.~\ref{fig1}b].
We focus on the state generated at intermediate values of the detuning during this preparation protocol. Similar to the previous section, we will use the total sweep time $T$ as a parameter to interpolate from a sudden quench to a perfectly adiabatic dynamics where the system is in the instantaneous ground state.
To describe the state of the system during this dynamics we introduce the following variational ansatz
\begin{align}
\nonumber \ket{ \phi(z_1,z_2) } = \mathcal{N} \, \mathcal{P} \, \Bigg[\bigotimes_{i=1}^N \left( 1+z_2 \sigma_i^+ \right)   \big( 1+ & z_1 \sigma_i^- \big) \Bigg]   \ket{\mathrm{RVB}},  \\
& z_1,z_2 \in \mathbb{C} \, ,
\label{ansatz}
\end{align}
where $\mathcal{P}$ is the projector on the sector of the Hilbert space that satisfies the blockade constraint, $\sigma^-_i=\ket{g}_i\!\bra{r}$, $\sigma^+_i=\ket{r}_i\!\bra{g}$ and $\mathcal{N}$ is a normalization constant. We allow the two variational parameters $z_1$ and $z_2$ to be complex to capture relative phases between fixed density subspaces. To understand this manifold of states it is instructive to consider the limiting cases. In one limit, where $z_1 = \infty$ and $z_2 = 0$, the state reduces to the trivial vacuum state, i.e., the initial state of the experimental state preparation protocol without any Rydberg excitation. 
In another limit, when $z_1 = z_2 = 0$, the state is simply the RVB state. In the vicinity of this point, the parameters $z_1$ and $z_2$ control the properties of the defects on top of the RVB state. These defects are monomers, i.e. vertices of the kagome lattice that are not covered by a dimer. Specifically, a finite value of the parameter $z_1$, results in the creation of nearest-neighbor monomer pairs that are created by removing a dimer from a dimer covering. The value of $z_1$ controls the density of such pairs. A finite value of $z_2$ effectively allows these monomer pairs to separate, introducing monomer pairs with larger intra-pair distances. Finally, in the limit when $z_1 = \infty$, all monomers are uncorrelated and their density is set by $z_2$. This last limiting case has been previously employed in variational studies of groundstates for Rydberg atom arrays on the square lattice, and its norm maps to a classical partition function with local weights~\cite{ji2011}.
An important feature of the state $\ket{\phi(z_1,z_2)}$ is, that it is a TN state of bond dimension 4 for all $z_1,z_2$. This follows from the observations that the RVB state is a TN state of bond dimension 2, and $\mathcal{P}$ is a TN operator with the same bond dimension. We note that expectation values for this state can be computed exactly by the contraction of a tensor network of bond dimension 8 (instead of the naively expected $4^2$)~\cite{Note1}.\\
\begin{figure}
\hspace*{-0.2cm}
\vspace*{-0.2cm}
\includegraphics[scale=0.48]{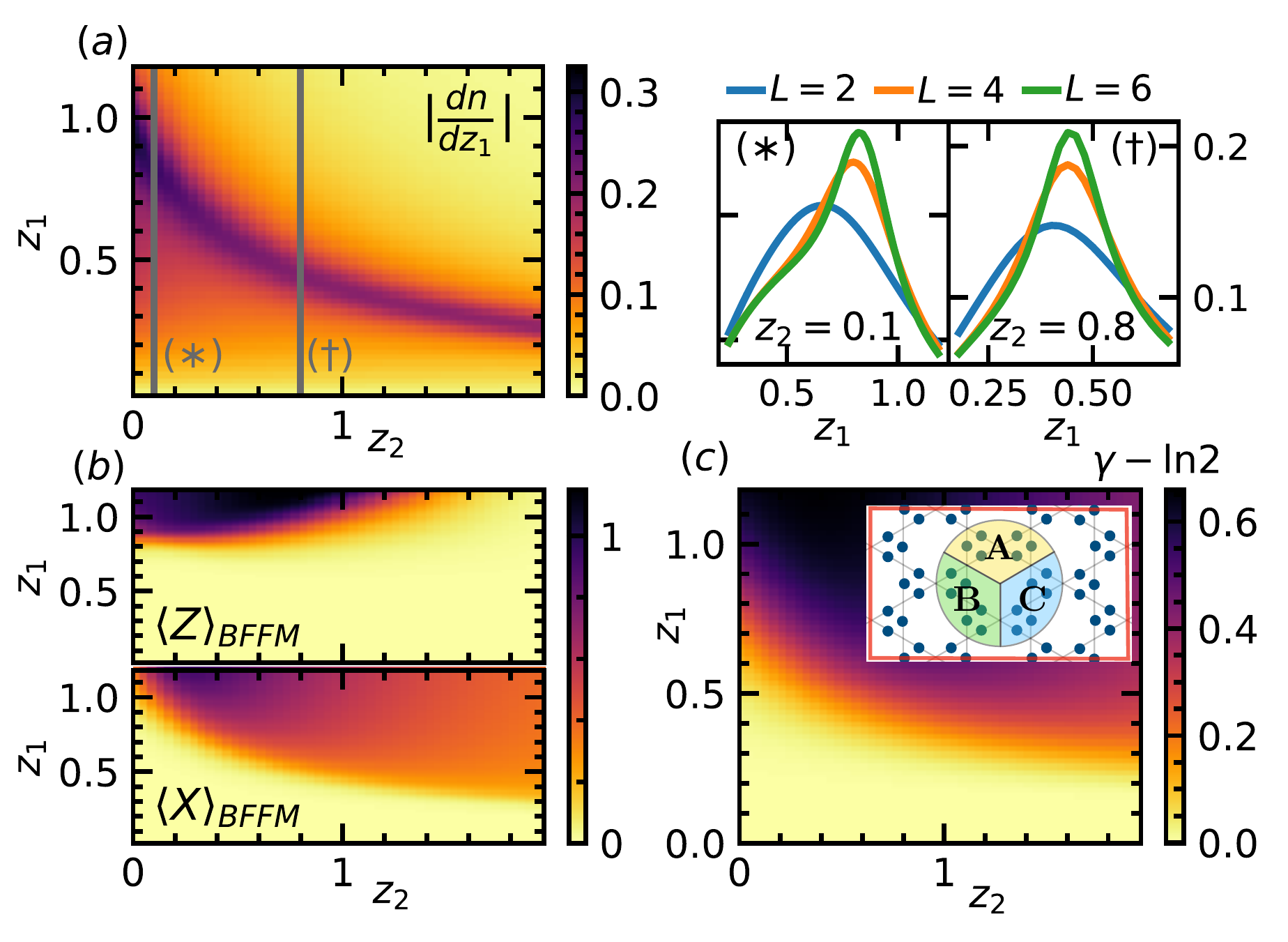}
\caption{(a) Derivative of the density w.r.t. the parameter $z_1$, computed from the TN representation of the ansatz state in Eq.~\eqref{ansatz} on an infinite cylinder of circumference $L=6$ tensors. On the right are plotted fixed $z_2$ section of the $(z_1,z_2)$ color plot for $z_2 = 0.1,0.8$. The scaling of the peak signals a continuous phase transition separating a topologically ordered phase from a trivial phase. (b) Diagonal and off-diagonal BFFM order parameters on the infinite cylinder of circumference $L=6$. (c) Topological entanglement entropy $\gamma = S_{AB} + S_{BC} + S_{AC} - S_A - S_B - S_C - S_{ABC}$ of the ansatz state computed on the finite cluster of $N=48$ sites depicted in the inset.}
\label{fig3}
\end{figure}
\indent
We demonstrate the effectiveness of the ansatz state Eq.~\eqref{ansatz} in Fig.~\ref{fig2}a, where we plot the optimized overlap with the dynamically evolving state for various total sweep times $T$ (solid lines) in a periodic cluster of $N=36$ atoms~\cite{Note1}. The dashed line denotes the optimized overlap with the instantaneous groundstate, which corresponds to a fully adiabatic sweep with $T=\infty$. The shaded red region indicates the topological phase arising in the ground state phase diagram. Our ansatz best describes the ground state in a neighborhood of the transition point between topological and disordered phases, at $\Delta/\Omega \simeq 1.4$. Remarkably, the highest overlaps with the variational state are not obtained for perfectly adiabatic sweeps, but instead for finite-time sweeps. Fig.~\ref{fig2}b shows that the fidelity slowly decreases with the number of atoms, but remains impressively large for all the system sizes considered ($>0.99$ for $N=48$ atoms). Similar to what we observed for the pure-RVB preparation protocol, the total sweep time for which maximal fidelities are reached increases linearly with $N$. In Fig.~\ref{fig2}c we plot the magnitude of the optimal values for the two variational parameters, for the fully adiabatic sweep (dashed line) and the optimal sweep rate for $N=36$.\\
These optimal values are to be located in the state phase diagram reported in Fig.~\ref{fig3}a that shows the derivative of the density of Rydberg excitations computed via TN methods on an infinite cylinder with circumference of length $12$ links of the kagome lattice ($L=6$ tensors)~\cite{Note1}. The presence of a peak in the derivative points at two distinct phases: an RVB-like phase and a trivial phase when $z_1,z_2$ are small and large respectively. A closer inspection of the scaling of the peak [Fig.~\ref{fig3}a, right] with the length of the circumference confirms a continuous phase transition separating a topological phase, connected to the RVB state and a trivial phase connected to the vacuum. We corroborate the topological nature of the former by computing the Bricmont, Fr\"olich, Fredenhagen, Marcu (BFFM)~\cite{bricmont1983,fredenhagen1983}, as defined in Ref.~\cite{Verresen2021}, and the topological entanglement entropy~\cite{Kitaev2003,levin2006}.
In Fig.~\ref{fig3}b we plot the diagonal and off-diagonal BFFM order parameters obtained from hexagonal loops of perimeter 18 links of the kagome lattice on an infinite cylinder with $L=6$~\cite{Note1}. The region where both these observables are small coincides with the conjectured topological phase, and we checked that, in this region, they vanish exponentially with increasing loop length~\cite{Note1}. 
We show in Fig.~\ref{fig3}c the topological entanglement entropy $\gamma$ of the state Eq.~\eqref{ansatz} on a periodic cluster of $N=48$ sites, obtained from $\gamma = S_{AB} + S_{BC} + S_{AC} - S_A - S_B - S_C - S_{ABC}$~\cite{Kitaev2006}, where $S_A$ is the entanglement entropy of the bipartition $A,A^c$ and the subsystems $A,B,C$ are depicted in the inset of Fig.~\ref{fig3}c. A value close to $\ln 2$ signals the emergence of $Z_2$ topological order in the region conntected to the RVB point.

\begin{figure}
\hspace*{-0.3cm}
 \includegraphics[scale=0.49]{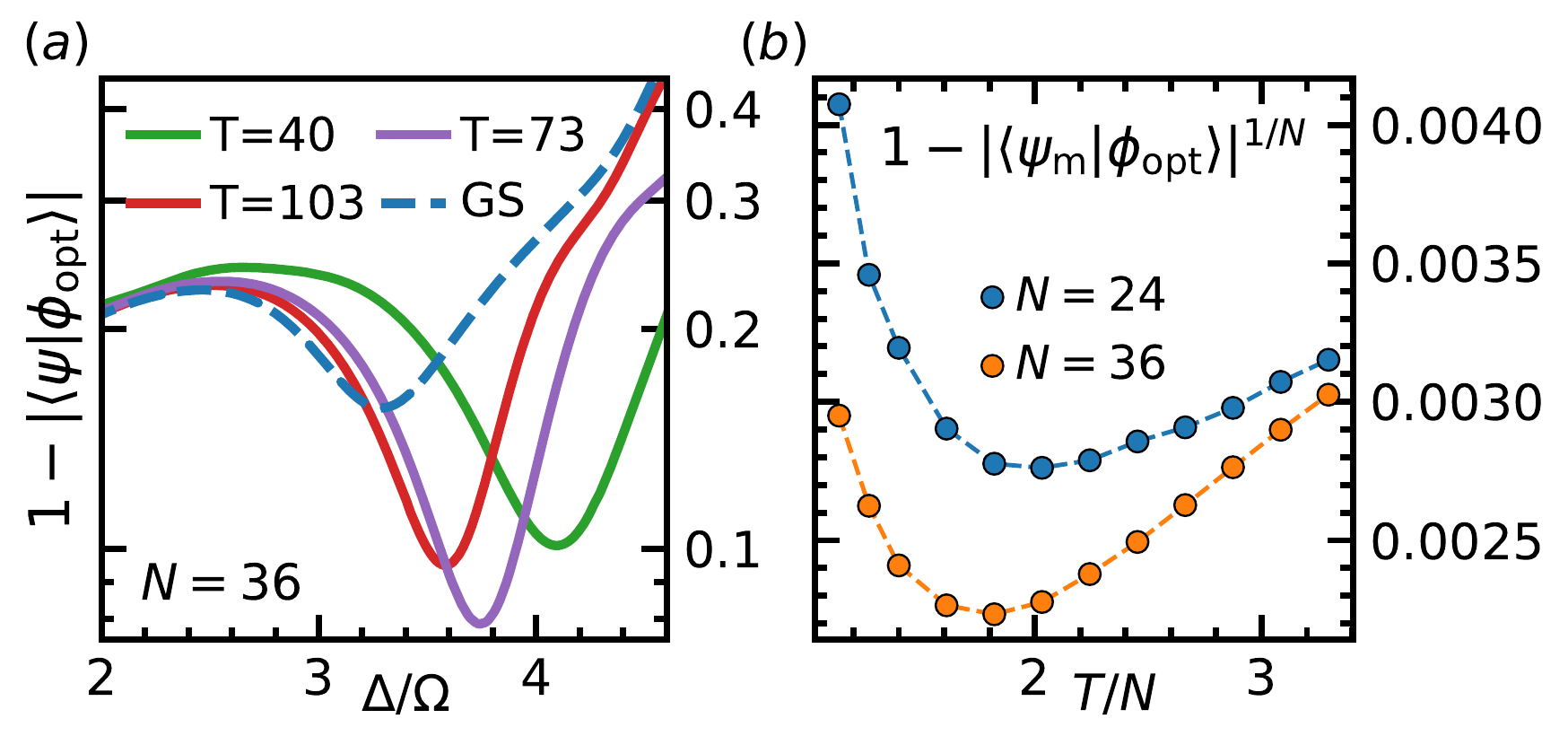}
\caption{(a) Optimized overlap between the ansatz Eq.~\eqref{ansatz} (without the projector $\mathcal{P}$) and the state prepared through the sweep depicted in the inset of Fig.~\ref{fig2}a for different total sweep times $T$, when the dynamics is generated by the full Rydberg Hamiltonian Eq.~\eqref{Ham_constr} with $R_b=2.4$. (b) Maximal fidelity per site obtained during the sweep as a function of total sweep time $T$ rescaled by the number of atoms $N$.  
}
 \label{fig4}
\end{figure}

\paragraph{Effect of long-range interactions. --} 
To study the state preparation dynamics generated by the full Rydberg Hamiltonian Eq.~\eqref{Ham_constr} we include long-range tails of the Van der Waals interaction and set $R_b = 2.4$. The maximal interaction distance between two excited Rydberg states is $|i-j| = \sqrt{13}$ in the length units defined previously. Moreover, we relax the radius of the hard constraint of Fig~\ref{fig1}a to half the length of a link of the kagome lattice, such that each triangle has at most one dimer. For comparison, we refer to the simplified model discussed in the previous sections as PXP model.
While the relaxation of the constraint notably improves the overlap with the RVB state~\cite{Note1}, the inclusion of long-range tails produces energy splittings in the fully-packed dimer coverings subspace of the diagonal part of the Rydberg Hamiltonian~\footnote{Interestingly, the fully-packed dimer configurations remain exactly degenerate upon inclusion of the shortest interaction beyond the blockade.}.
This fact generates a complex pattern of phases between the maximal-density components of the prepared state, yielding suppressed overlaps between the RVB state and the state reached at the end of the sweep in Fig.~\ref{fig1}a. We refer to \cite{Note1} for a more detailed discussion of this issue, and we stress here that high RVB fidelities are obtained when only the absolute value of the components of the prepared state is considered. Gaining a deeper understanding of the effect of these phases and their control is crucial for the experimental applicability of the preparation protocol discussed in the first part of this Letter.\\
\indent
We now focus on the ansatz Eq.~\eqref{ansatz} and its capability to characterize the state dynamically prepared at finite $\Delta/\Omega$. We note that the constraint relaxation produces a non-negligible projection of the latter on the subspace violating the dimer constraint~\cite{Semeghini2021}. When projecting out this component, and upon proper renormalization, Eq.~\eqref{ansatz} yields fidelities comparable to the PXP model~\cite{Note1}. However, in an attempt to capture this constraint-violating component, we remove the projector $\mathcal{P}$ in the overlap optimization. The resulting state-phase diagram is qualitatively unchanged w.r.t. the one in Fig.~\ref{fig3}~\cite{Note1}. We plot in Fig.~\ref{fig4}a the outcome of the optimization for a periodic cluster of $N=36$ atoms, for various total sweep times $T$ (solid lines), and including the fully adiabatic sweep (dashed line). The result is analogous to what is depicted in Fig.~\ref{fig2}a for the PXP model, where the smaller maximal overlaps are to be compared with the much larger Hilbert space dimension: $2^{24}$ vs $\simeq 2^{17}$ for $N=36$. Remarkably, the fidelity per site increases with increasing $N$ for the two system sizes considered, as reported in Fig.~\ref{fig4}b.

\paragraph{Outlook. --} 
We discussed dynamical preparation of  topological spin liquids in Rydberg atom arrays. First we showed that the pure RVB state can be reached with impressively high fidelity in a time that scales linearly with the number of atoms.
We then showed that the non-equilibrium state observed in~\cite{Semeghini2021} is well described by a two-parameter family of TN states with small bond dimension. The latter includes the topologically ordered RVB state and the vacuum. We exploit this TN representation to study the properties of the prepared state on unprecedentedly large systems, and infer about the stability of topological order in the thermodynamic limit. We find  that our ansatz is fully consistent with a topological spin liquid in a finite region in parameter space.
Our work clarifies the nature of non-equilibrium state experimentally prepared in \cite{Semeghini2021}, and provides the tools for performing large-scale classical simulations that might serve as guidance for probing topological quantum matter in future  quantum simulator experiments. These studies can be extended along several directions. For instance, our approach can be used to explore non-trivial dynamics of anyonic excitation as well as effects associated with errors due to e.g. spontaneous emissions. Moreover, the dynamical preparation of the pure RVB state is not limited to the ruby lattice described in the present work. In particular, one can explore if this method can applied to other systems with a ground state degeneracy growing exponentially with the number of atoms (see e.g. Ref.~\cite{Samajdar2021}). In such systems dynamical preparation protocol can be potentially used to engineer other kinds of exotic phases of matter in a wide variety of lattice geometries.

\paragraph{Acknowledgements. --} We acknowledge useful discussions with  I. Cong, G. Giudice, N. Maskara, S. Sachdev, R. Samajdar, G. Semeghini, R. Verresen, A. Vishwanath, and T. Zache. This research was supported by the Army Research Office (Grant Number W911NF-21-1-0367),  a discovery grant by the Erwin Schr\"odinger Center for Quantum Science, CUA, NSF and DOE.
GG acknowledges support from the Deutsche Forschungsgemeinschaft (DFG, German Research Foundation) under Germany's Excellence Strategy -- EXC-2111 -- 390814868 and from the ERC grant QSIMCORR, ERC-2018-COG, No. 771891. \\
{\it Note added}: while completing this manuscript we became aware of a related variational study of non-equilibrium topological state preparation Ref.~\cite{cheng2021}.

\bibliography{tRVB.bib}

\appendix
\clearpage
\newpage

\renewcommand{\theequation}{S\arabic{equation}}
\renewcommand{\thefigure}{S\arabic{figure}}
\setcounter{figure}{0}

\onecolumngrid
	
\begin{center}
	\textbf{\large Supplementary Material}
\end{center}
\vspace*{0.5cm}
\twocolumngrid

\begin{figure}
 \hspace*{-0.5cm}
 \includegraphics[scale=0.51]{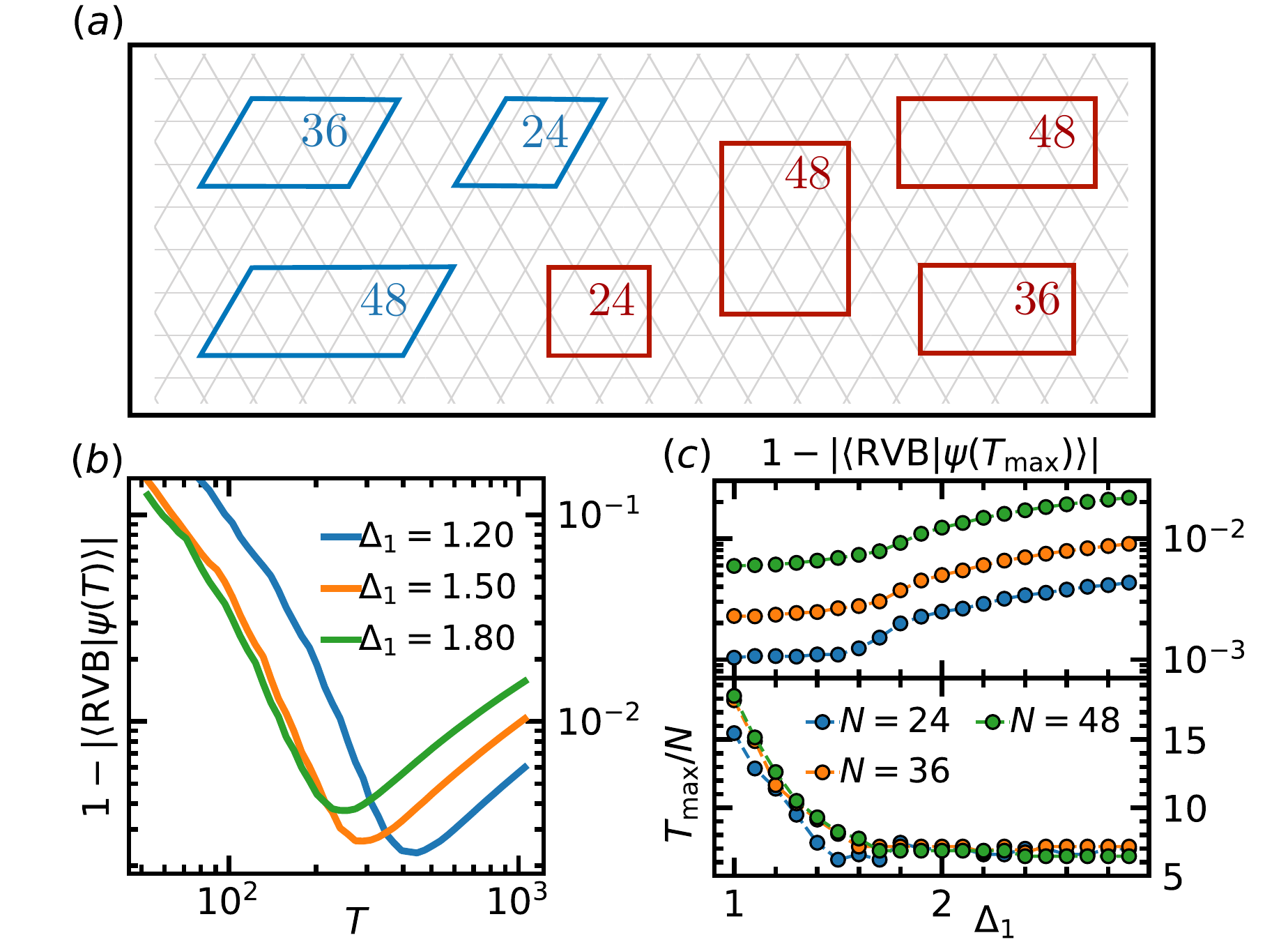}
\vspace*{-0.4cm}
\caption{(a) Periodic clusters employed for the numerical calculations of the exact dynamics. (b) Overlap between the prepared state and the RVB state for different values of the final detuning $\Delta_1$, the other sweep parameters being fixed as in the main text. (c) Overlap between the prepared state and the RVB state maximized over $T$ as a function of $\Delta_1$. (d) Total sweep time $T_{\mathrm{max}}$ for which the maximum RVB overlap is obtained as a function of the final detuning $\Delta_1$. The y-axis is rescaled by the number of sites, showing almost perfect collapse for all considered values of $\Delta_1$. }
\label{fig1_SI}
\end{figure}

\paragraph{Exact diagonalization and protocol optimization -- } We detail here the methods employed for the computation of the exact dynamics, the choice of the optimal sweep parameters, and the state optimization.\\
Fig.~\ref{fig1_SI}a shows all the periodic clusters where numerical simulations have been performed. The data presented in Fig.~1d and Fig.~2 of the main text are obtained from the blue clusters on the left of the panel, the groundstate fidelity susceptibility and RVB overlap in Fig.~1c(main text) from the red 48-atoms cluster on the top right, the entanglement entropy plotted in Fig.~3c(main text) from the other red 48-cluster on the right of the panel. The analysis performed in the main text has been carried out on all clusters depicted in Fig.~\ref{fig1_SI}a and all stated results are unaffected by the aforementioned choices. The simulated dynamics respects translation and point group symmetries of each cluster, therefore numerical calculations have been restricted to the most symmetric block of the Hamiltonian.\\
\indent
The preparation protocol depicted in Fig.~1a(main text) consists of a first ramp to turn on the Rabi frequency at constant detuning $\Delta_0$, followed by a second ramp to turn on the detuning at constant $\Omega$ to a final value $\Delta_1$, and by a final ramp to switch off $\Omega$ at constant detuning $\Delta_1$. The durations of the three ramps and the total duration of the sweep are $T_1,T_2,T_3,T$ respectively. The ramps have been smoothed via a uniform filter to eliminate non-adiabatic effects on short time scales. The initial value of the detuning $\Delta_0$ is irrelevant, as the initial state (the vacuum) is the groundstate of the instantaneous Hamiltonian at $t=0$ as long as $\Delta_0 < 0$. The parameter $T_1/T,T_2/T,T_3/T$ have been fixed to give the highest fidelity between the prepared state and the RVB state, for fixed system size and sweep total duration $T$, and it has been verified that the result is independent of $T$. We cannot exclude that changing the shape of the functions $\Delta(t)$ and $\Omega(t)$ might lead to improved RVB fidelities, but we find a qualitative difference in our results unlikely. Analogous considerations apply to the protocol depicted in the inset of Fig.~2a(main text).\\
Fig.~\ref{fig1_SI}b shows the RVB overlap of the prepared state as a function of the total sweep time, for different values of the final detuning $\Delta_1$. Both the optimal overlap and the sweep rate at which the optimum is achieved dislpay a clear dependence on $\Delta_1$. Optimal overlaps and the corresponding sweep rates are plotted in Fig.~\ref{fig1_SI}c and Fig.~\ref{fig1_SI}d, respectively. For $\Delta_1 \gtrsim 1.6$ the optimal total time $T_{\mathrm{max}}$ saturates, while the optimal overlap decreases with increasing $\Delta_1$. $\Delta = 1.6$ is remarkably close to the phase transition occurring from the disordered to the RVB-like phase in the groundstate [see Fig.~1c(main text)]. With decreasing $\Delta_1$ and $\Delta_1 \lesssim 1.4$ the optimal overlap ceases to increase significantly, while $T_{\mathrm{max}}$ increases. We choose the optimal sweep rate as the one that minimizes $T_{\mathrm{max}}$ and maximizes the overlap at the same time. Finally, as for the system size dependence of the optimal time, Fig.~\ref{fig1_SI}c clearly demonstrates a linear increase of $T_{\mathrm{max}}$ with the number of atoms independently of $\Delta_1$. \\
The optimization of the overlap between the ansatz in Eq.~(2)(main text) text and the state during the preparation dynamics has been carried out by direct evaluation of the overlap, using the Nelder-Mead method implemented in the python library SciPy.

\begin{figure}
 \hspace*{-0.4cm}
 \includegraphics[scale=0.49]{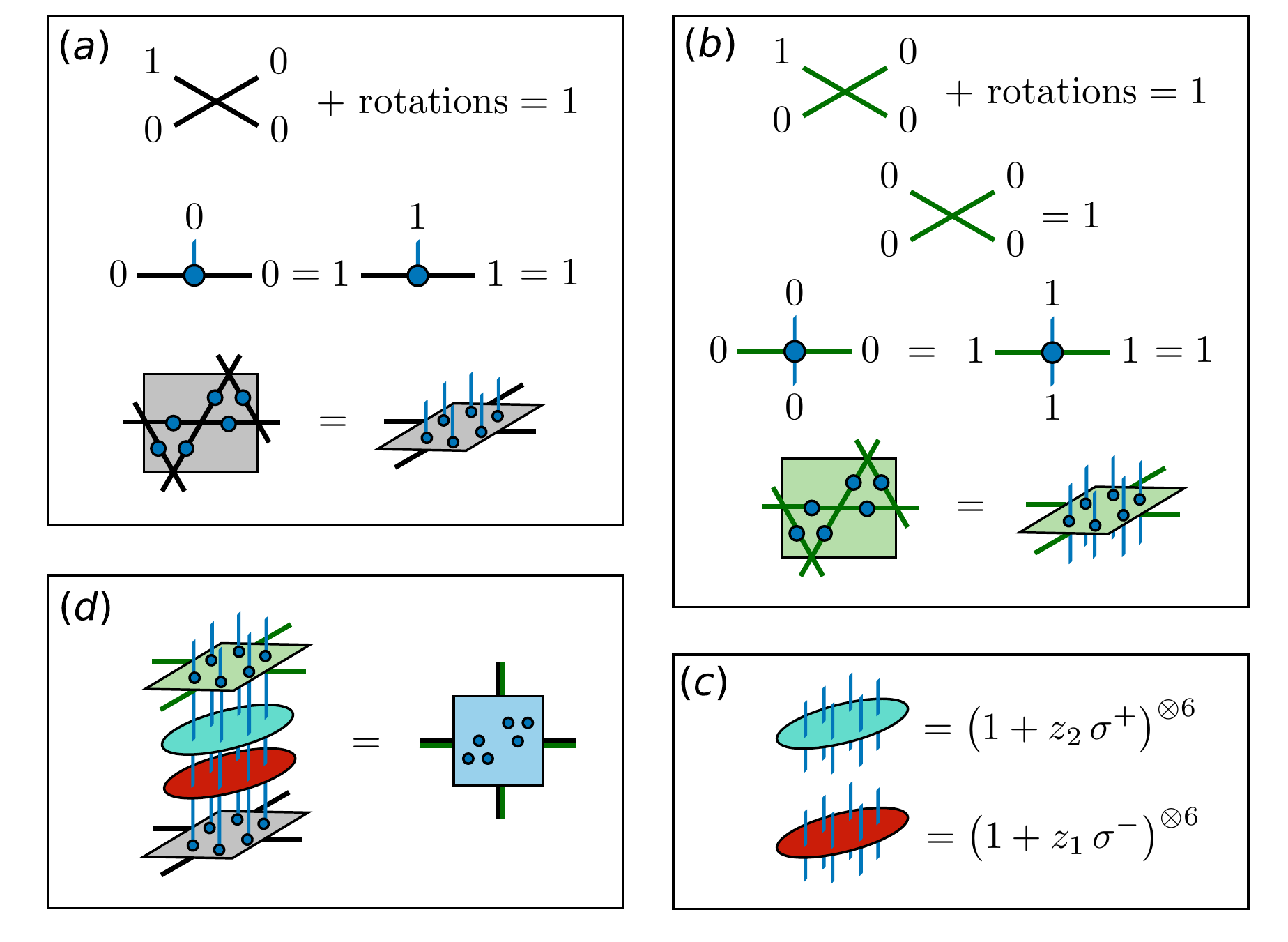}
\vspace*{-0.4cm}
\caption{Tensor network representation of the ansatz Eq.~(2)(main text). (a) The TN for the RVB state (grey tensor) can be obtained by contracting the 4-legs vertex tensor with the 3-legs projector. The latter maps the virtual spaces of bond dimension $2$ (black lines) to the physical space of a unit cell of the ruby lattice (blue lines). The virtual state $1$ signals the presence of a dimer on the corresponding link of the kagome lattice, and it is diagonally mapped to the physical space via the 3-legs projector. (b) The projector on the constrained Hilbert space of the Rydberg atoms array (grey tensor) can be expressed as a TN operator of bond dimension $2$ by adding a component to the RVB vertex tensor that allows for the absence of a dimer entering a vertex and a physical leg to the diagonal map from virtual to physical space. (c) The two local operators containing the variational parameters $z_1$ and $z_2$ are trivially represented as TN operators of bond dimension $1$. (d) Applying the two local operators (c) to the RVB state (a), and projecting out the components that do not satisfy the Rydberg constraint results in a TN state of bond dimension $4$ (light blue tensor).  }
\label{fig2_SI}
\end{figure}

\begin{figure}
 \hspace*{-0.25cm}
 \includegraphics[scale=0.48]{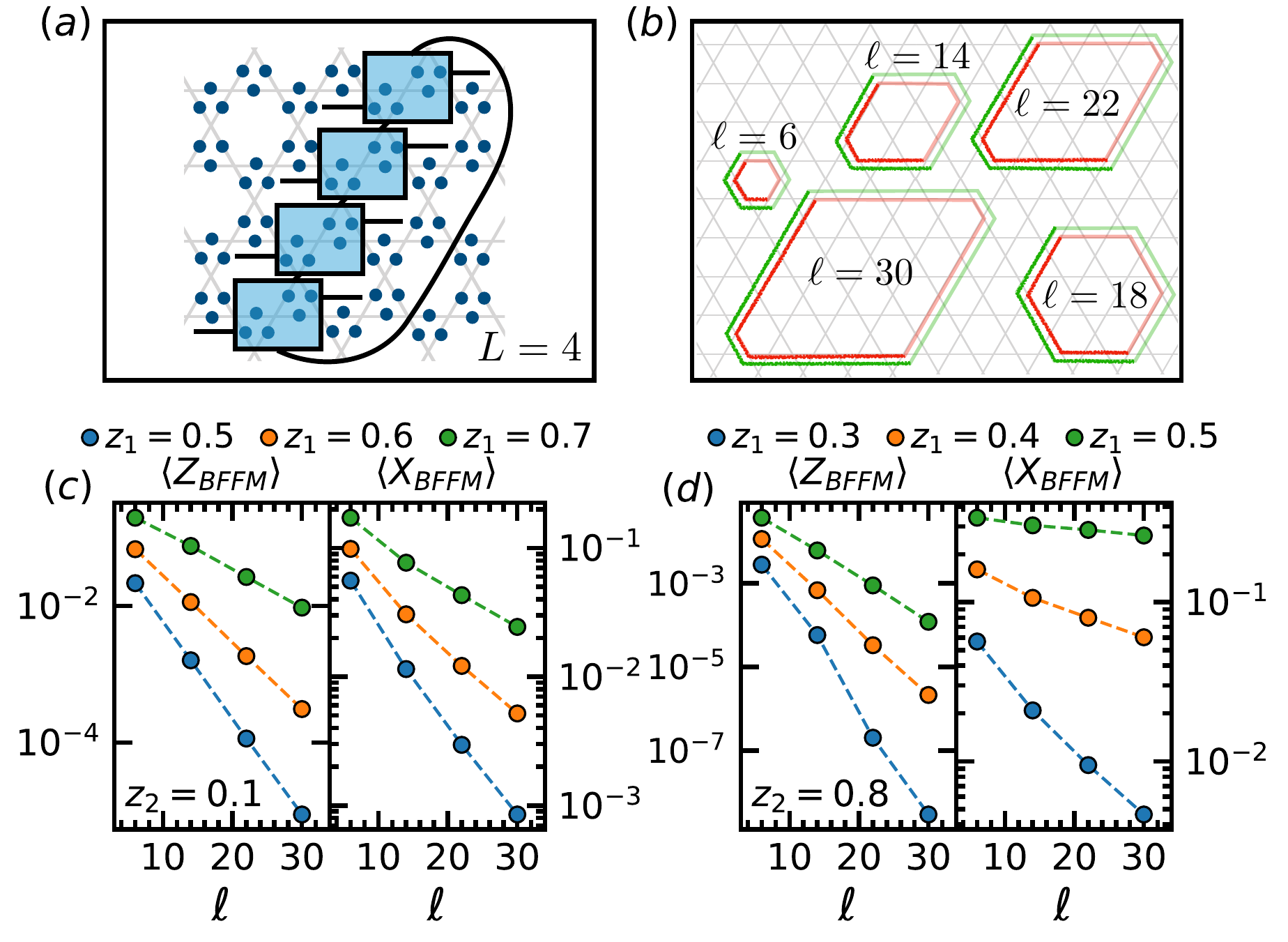}
\vspace*{-0.3cm}
\caption{(a) Geometry of the tensor network representation of the ansatz Eq.~(2)(main text). Each tensor cover a unit cell of the ruby lattice, therefore including 6 atoms per tensor. The picture shows a network of $L=4$ tensors wrapped around a cylinder. Infinite cylinder calculations can be easily carried out for $L \le 6$. (b) Some of the loops used for the computation of the BFFM order parameters from the infinite cylinder transfer matrix, where $\ell$ denotes the length of the perimeter. Green and red loops are used for diagonal $Z$ loops and off-diagonal $X$ loops respectively. (c-d) Scaling with $\ell$ of the BFFM order parameters for parallelogram-shaped loops like the $\ell=30$ loop of panel (b). For $z_2=0.1$ both $Z_{\mathrm{BFFM}}$ and $X_{\mathrm{BFFM}}$ vanish exponentially for all $z_1$ considered. For $z_2=0.8$, $X_{\mathrm{BFFM}}$ exhibits exponential scaling only for sufficiently small $z_1$, while $Z_{\mathrm{BFFM}}$ appears to scale exponentially for all $z_1$ considered, even outside the topologically ordered phase. }
\label{fig3_SI}
\end{figure}

\paragraph{TN representation and cylinder transfer matrix --} As pointed out in the main text, the ansatz [Eq.~(2)(main text)] can be written as a tensor network state of bond dimension 4 and it is represented in Fig.~\ref{fig2_SI}. The RVB state is a TN state of bond dimension 2 that can be easily constructed by contracting vertex tensors that enforce the presence of a single dimer entering each vertex [Fig.~\ref{fig2_SI}a]. The two variational parameters $z_1$ and $z_2$ are encoded in the action of local operators on the RVB state [Fig.~\ref{fig2_SI}c]. The latter do not increase the bond dimension of the tensors. The result is a TN state that includes a component that does not satisfy the dimer constraint. We project out this component with a TN operator of bond dimension 2 depicted in Fig.~\ref{fig2_SI}b, making the ansatz a TN state of bond dimension $D=4$. \\
\indent With the TN representation at hand the computation of expectation values on the state can be carried out in two steps. First the cylinder transfer matrix is constructed by contracting bra and ket layers of the state tensor (the double tensor) along a circumference [see Fig.~\ref{fig3_SI}a]. Each virtual bond of the transfer matrix has dimension $D^2 = 16$, however, a factor of $2$ can be easily eliminated by noticing that, since a projector squares to itself, only one projector layer is necessary when computing the double tensor.
The (left and right) maximal eigenvalue-eigenvectors of the cylinder transfer matrix are then obtained, either from Lanczos iterations or power method. The dominant eigenvectors are then used to sandwich an arbitrary number of cylinder transfer matrices with local operator inserted between bra and ket in the double tensors [see e.g.~\cite{Zauner2015} for a more detailed illustration]. We employed this technique to compute the local density and the BFFM order parameters plotted, respectively, in Fig.~3a and Fig.~3b of the main text. We refer to \cite{Verresen2021} for the definition of these operators. In Fig~\ref{fig3_SI}b we show some of the loops used for the computation of the BFFM order parameters, in particular, the $\ell=18$ hexagonal loop is the one employed in Fig.~2b(main text). Fig.~\ref{fig3_SI}c and Fig.~\ref{fig3_SI}d show the scaling with the loop length of the BFFM order parameters. In a topologically ordered phase they should vanish exponentially, while they should approach a constant value in a trivial phase. Fig~\ref{fig3_SI}c shows a clear exponential scaling when $z_2 = 0.1$ for different values of $z_1$ within the topological phase, that occurs when $z_1 \lsim 1$ [see the peak position in the density derivative in Fig.~3a(main text)]. In Fig~\ref{fig3_SI}d we plot the same quantities for $z_2 = 0.8$. While the decay length of the off-diagonal order parameter vanishes as the boundary of the topological region is approached, the diagonal order parameter seems to scale exponentially even outside this region. These observations imply deconfinement of both electric and magnetic excitations in the topologically ordered phase.

\begin{figure}
 \hspace*{-0.2cm}
 \includegraphics[scale=0.5]{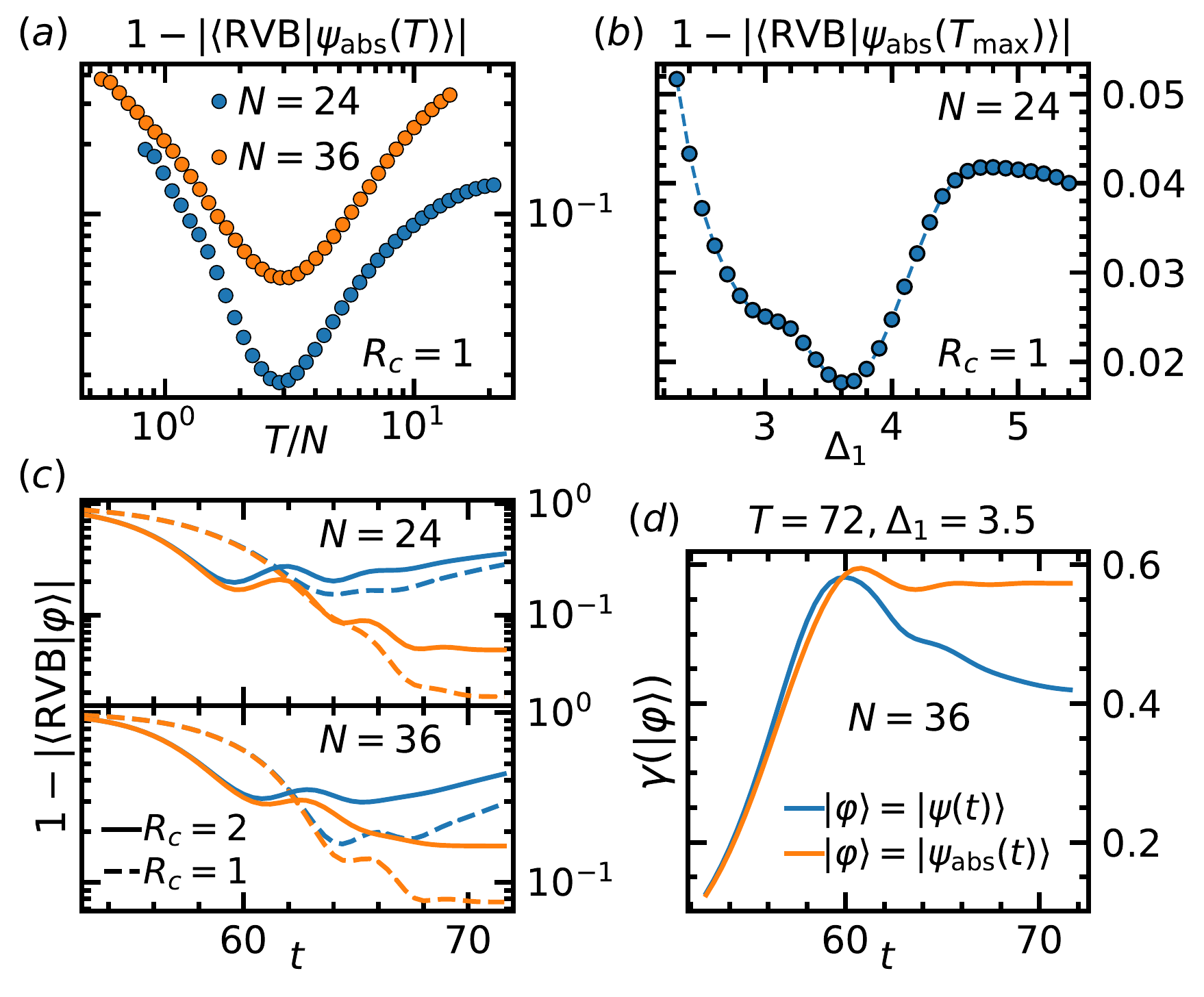}
\caption{(a) Overlap between the absolute value of the prepared state and the RVB state, with a constraint radius $R_c=1$ and blockade radius $R_b = 2.4$. (b) Overlap between the absolute value of the prepared state and the RVB state maximized over $T$ as a function of $\Delta_1$, with $R_c=1$ and $R_b=2.4$. (c) Comparison between different constraint radius for $T=72$ and $\Delta_1 = 3.5$. Dashed and solid lines represent $R_c=1$ [at most one Rydberg state per triangle of the kagome lattice] and $R_c=2$ [see Fig.~1a(main text)]. Relaxing the constraint increases the RVB fidelity reached in the final state, both for the prepared state (blue) and its absolute value (orange). (d) Topological entanglement entropy [see Fig.~3c(main text)] of the prepared state (blue) vs its absolute value (orange).}
\label{fig4_SI}
\end{figure}

\begin{figure}
 \hspace*{-0.3cm}
 \includegraphics[scale=0.5]{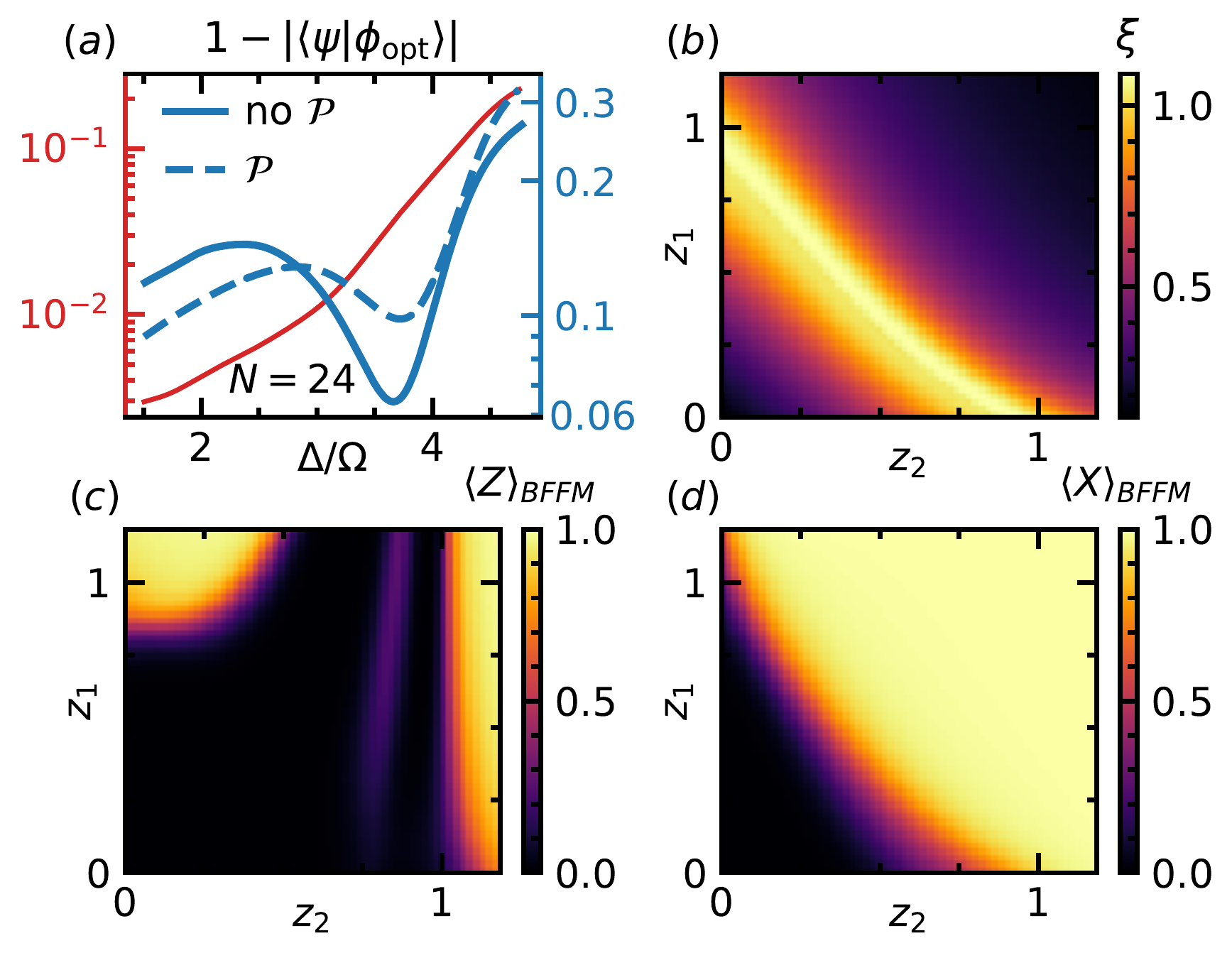}
\vspace*{-0.3cm}
\caption{(a) Comparison between the optimized overlap obtained from the dynamical preparation with the full Rydberg Hamiltonian when the variational ansatz is taken as the state Eq.~(2)(main text) with the projector $\mathcal{P}$ (dashed line) and without the projector $\mathcal{P}$ (solid line). The solid red line is the optimized overlap when the dynamically prepared state is projected onto the $R_c=2$ subspace and renormalized to unity. (b) Correlation length $\xi = 1/\log ( \lambda_0 / \lambda_1 )$ computed from the largest non-degenerate eigenvalues $\lambda_0 > \lambda_1$ of the cylinder transfer matrix, for the TN state in Fig.~\ref{fig2_SI} when the projector $\mathcal{P}$ is removed. (c-d) Diagonal and off-diagonal BFFM order parameters computed on the infinite cylinder with $L=8$, on hexagonal loop with $\ell = 18$ depicted in Fig.~\ref{fig3_SI}b). }
\label{fig5_SI}
\end{figure}

\paragraph{Constraint relaxation and long-range interactions -- }
\hspace{-1mm}In the main text, we briefly discussed the effect of the relaxation of the constraint in Fig.~1a(main text) and the inclusion of longer range tails of the Van der Waals potential on the protocol for the preparation of pure RVB states. We claimed that, while the former enhances RVB overlaps of the prepared state, the latter produces a complex pattern of phases between different components within the maximal density subspace that leads to a suppression of RVB fidelities. Here we show that this suppression is not present when these phases are neglected by taking the absolute value of the prepared state.
Numerical calculations are performed on the blue clusters of Fig.~\ref{fig1_SI}a, for which the maximal interaction distance is $\sqrt{13}$ in units of the minimum distance between the atoms. In what follows we define $R_c$ as the constraint radius: states where more than one Rydberg is excited in a circle of radius $R_c$ are discarded from the Hilbert space. $R_c = 1$ thus corresponds to at most one excited state per kagome lattice triangle, while $R_c = 2$ is equivalent to the exact dimer constraint as depicted in Fig.~1a(main text). Notice that $R_c = 1$ is an almost exact approximation of the full Rydberg Hamiltonian when the blockade radius is set to $R_b=2.4$, since $(R_b)^6 \simeq 191$. \\
\indent
We plot in Fig.~\ref{fig4_SI} RVB fidelities of the absolute value of the prepared state as a function of the total sweep time $T$ for different systems sizes [Fig.~\ref{fig4_SI}a] and as a function of $\Delta_1$ at the optimal sweep rate [Fig.~\ref{fig4_SI}b] for $R_c =1 $. In full analogy with the PXP model, the best RVB overlap is obtained at intermediate total sweep times that scale linearly with $N$. The value of $\Delta_1$ yielding the highest overlap with the RVB state is sharply around $\Delta_1 \simeq 3.5$. Therefore we chose $\Delta_1=3.5$ for the data presented in Fig.~\ref{fig4_SI}a.
In Fig.~\ref{fig4_SI}c we compare the overlap in the final part of the sweep between RVB state and the absolute value of the prepared state with $R_c = 1$ (orange dashed line) and $R_c = 2$ (orange solid line). The relaxation of the constraint clearly enhances RVB fidelities. In the same figure, we plot RVB overlaps when also phases in the prepared state are taken into account (blue lines). Independently of $R_c$, the phase pattern generated by the full Van der Waals potential lowers the overlap between the final state and the RVB state. We address the effect of this suppression on the topological properties of the prepared state $\ket{\psi}$ in Fig.~\ref{fig4_SI}d, where we plot the topological entanglement entropy $\gamma$ [see Fig.~3c(main text)] in the final part of the sweep, both for $\ket{\psi}$ and its absolute value $\ket{\psi_{\mathrm{abs}}}$. While $\gamma(\ket{\psi})$ reaches an approximate plateau (orange line), $\gamma(\ket{\psi_{\mathrm{abs}}})$ decreases in the final stage of sweep, where the detuning is kept constant to $\Delta_1$ and the Rabi frequency is switched off. The RVB overlap suppression thus also leads to topological entropy suppression. We leave open the problem of stabilizing an RVB-like state as a stationary state at the end of the sweep.\\
\indent
When we extended the ansatz Eq.~2(main text) to the dynamics generated by the full Rydberg Hamiltonian, we removed the projector $\mathcal{P}$ that enforces at most one dimer per vertex of the kagome lattice. We now show that this choice leads indeed to higher overlaps with the variational state during the dynamical preparation. In Fig.~\ref{fig5_SI}a we plot the optimized overlap with (dashed line) and without (solid line) the projector $\mathcal{P}$. While in the former case the maximum overlap is $\simeq 0.9$ for $N=24$ atoms, in the latter it reaches $0.94$ at intermediate $\Delta/\Omega$. We note that the projected ansatz well describes the prepared state through the whole preparation process, as demonstrated by the solid red line in Fig.~\ref{fig5_SI}a, that stands for the optimized overlap when the dynamically prepared state is projected onto the $R_c=2$ subspace and renormalized to unity.\\
\indent
The state-phase diagram originating from this different variational ansatz is qualitatively unchanged, as we demonstrate in Fig.~\ref{fig5_SI}. In Fig.~\ref{fig5_SI}b we plot the correlation length obtained from the two largerst non-degenerate eigenvalues of the cylinder transfer matrix $\lambda_0$ and $\lambda_1$ via the formula $\xi = 1/\log ( \lambda_0 / \lambda_1 )$~\cite{Zauner2015}. A clear peak separates the topologically ordered RVB point $z_1 = z_2 = 0$ from the trivial vacuum $z_1 = \infty, z_2 = 0$ and from the trivial completely filled state $z_1 = 0, z_2 = \infty$. Differently from the projected case, the topologically ordered region is bounded in parameter space, since a trivial state is reached in both the $z_1$ and $z_2$ directions. Moroever, in Fig.~\ref{fig5_SI}c and Fig.~\ref{fig5_SI}d we show that this bounded region is topologically ordered by plotting the diagonal and off-diagonal BFFM order parameters in the $(z_1,z_2)$-plane.
\end{document}